\documentstyle[12pt]{article}
\normalsize

\newcounter{sxn}

\newcounter{axn}

\def\br{\begin{thebibliography}{abc}}
\def\er{\end{thebibliography}}

\date{}

\begin{document}
\bibliographystyle{unsrt}
\footskip 1.0cm
\thispagestyle{empty}
\begin{flushright}
UMN-TH-1049/92\\
TPI-MINN-92/36-T\\
NSF-ITP-92-135i\\
September 1992\\
\end{flushright}
\vspace*{6mm}
\centerline {\Large VORTEX PRODUCTION IN A FIRST ORDER PHASE-TRANSITION}
\vspace*{3mm}
\centerline {\Large AT FINITE TEMPERATURE}
\vspace*{8mm}
\centerline {\large Sumantra Chakravarty}
\vspace*{5mm}
\centerline {\it School of Physics and Astronomy, University of Minnesota}
\centerline {\it Minneapolis, Minnesota 55455, USA}
\vspace*{8mm}
\centerline {\large Ajit Mohan Srivastava \footnote {\small Present address:
Institute for Theoretical Physics, University of
California, Santa Barbara, CA 93106, USA.} }
\vspace*{5mm}
\centerline {\it Theoretical Physics Institute, University of Minnesota}
\centerline {\it Minneapolis, Minnesota 55455, USA}

\vspace*{5mm}

\baselineskip=18pt

\centerline {\bf ABSTRACT}
\vspace*{3mm}
We simulate the production of vortices in a first order phase transition
at finite temperature. The transition is carried out by randomly nucleating
critical bubbles and the effects of thermal fluctuations (which could be
relevant for vortex production) are represented by randomly nucleating
subcritical bubbles.  Our results show that the presence of subcritical
bubbles suppresses vortices with clear and prominent profiles, though
net number of vortices is consistent with theoretical estimates. We also
determine the typical  speed of vortices arising due to randomness
associated with the phase transition to be about 0.5.

\newpage

\newcommand{\be}{\begin{equation}}
\newcommand{\ee}{\end{equation}}

\baselineskip=18pt
\setcounter{page}{2}

\vskip .1in

The study of a  system undergoing a phase transition  can have many
important implications; production of topological defects being one of
them [1]. An effective theory  describing the
transition can be constructed by considering the evolution of its
order-parameter field ($\Phi$). Above the transition temperature
$\Phi$ is zero whereas below the transition temperature $\Phi$ assumes non-zero
vacuum expectation value. For the case of first order phase transition,
bubbles of lower temperature phase nucleate within the supercooled
(metastable) high temperature phase as the temperature is lowered through
the transition temperature. Associated with the production of such a true
vaccum bubble there is a gain in the volume energy and a loss in the surface
energy. There is thus a critical size for which the bubble formation is
energetically favored. These critical bubbles expand and coalesce with one
another to fill the space with the lower temperature phase [2].
Though the energetically unfavorable sub-critical bubbles collapse eventually,
they survive long enough and may significantly affect the history
of the phase transition, especially the process of vortex formation (see [3]).

 A numerical simulation of the dynamical production of vortices through bubble
collisions
has been carried out in [3]. However, the probability of bubble
nucleation was chosen a priorie there as the intention was to check the
theoretical prediction of 1/4 vortices per bubble (for U(1) global strings).
Further, the bubbles which were randomly nucleated, were
all critical bubbles appropriate for the zero temperature case. In this paper
we extend the investigation of [3] by considering the bubbles at finite
temperature. For this case we make an estimate of the nucleation rate
(along with the pre-exponential factor) and nucleate bubbles with corresponding
probabilities. We carry out the simulation of the phase transition
for the case when a U(1) global symmetry is spontaneously broken. We consider
the nucleation of critical bubbles at finite temperature and include
subcritical bubbles as representing the dominant contribution of thermal
fluctuations in the process of vortex formation. Subcritical bubbles
have been considered as playing a crucial role in the phase transition
by Gleiser, Kolb,  and Watkins (see [4]). We would like to mention that we
do not attempt to incorporate all possible effects of thermal fluctuations.
Such effects could very well  affect our results on the estimation of the
random speeds of the vortices. However, we believe that an important class of
thermal fluctuations is represented by the subcritical bubbles, which we do
incorporate, especially
from the point of view of estimating the vortex formation probability.

In this work we consider a 2+1 dimensional case and adopt the following
Lagrangian:

\be {\sl L} =
         {1\over 2}\partial_\mu\Phi^\dagger\partial^\mu\Phi-V(\Phi ) \ee

   \be  V(\phi ) = {1\over 2}m^2\phi^2 -{1\over 3}\delta\phi^3
           +{1\over 4}\lambda\phi^4  \ee
where $\Phi =\phi e^{i\theta}$ is a complex scalar field.
The coefficients $m,~\delta$, and $\lambda$
in the effective potential $V(\phi )$ are assumed to be temperature dependent
renormalized quantities. Depending on the values of these parameters, the
potential has minima at
\be\phi=0 {\quad\rm~and~at}\quad \phi =\eta ={1\over 2\lambda}
      \left\{\delta +(\delta^2 -4\lambda m^2)^{1/2}\right\}\ee

If $V(\eta )<V(0)$, true ground state for the system is given by $\phi =\eta$
and the global $U(1)$ symmetry in (1) is
spontaneously broken. Topological defects in the form of global
vortices appear because of loops with non-trivial winding number in the
$U(1)$ vacuum manifold. If the system is supercooled and  still remains in
the false  vacuum $\phi =0$ below the transition temperature, it tunnels
through the barrier to the true vacuum. At zero temperature, the
tunneling probability can be calculated by finding the bounce solution
which is a solution of three dimensional Euclidean equations of motion [2].
However, at finite temperature, the theory becomes effectively 1+1
dimensional if the temperature is sufficiently high and the tunneling
probability is governed by the solutions of two dimensional Euclidean
equations of motion [5].

The tunneling probability per unit volume (area) per unit time in the high
temperature approximation is given by [5] (we use $\hbar =c=1$ in this work)

\be \Gamma = A~e^{-S_2(\phi )/T} \ee
where $S_2(\phi )$ is the two dimensional Euclidean action for a field
configuration that satisfies the classical Euclidean equations of motion.
The dominant contribution to $\Gamma$ comes from the least action
$O(2)$ symmetric configuration which is a solution of the following equation.

\be{d^2\phi\over dr^2} +{1\over r}{d\phi\over dr} -V^{\prime}(\phi )=0\ee

\noindent where $r \equiv r_E = \sqrt{{\vec x}^2 + t_E^2}$, subscript E
denoting the coordinates in the Euclidean space.

The boundary condition imposed on $\phi$  are

 $$ \phi =0 \qquad r\rightarrow\infty $$
 \be {d\phi\over dr}=0 \qquad r=0 \ee

 Bounce solution of Eq.(5) can be analytically obtained in the `thin wall'
limit where

\be\epsilon =V(0)-V(\eta )\ee

\noindent is much smaller
than the barrier height. However, such bubbles are large and
not suitable for numerical simulation. We therefore work with thick wall
bubbles and choose parameters such that the bubble size is small. The bubble
profile then has to be obtained by numerically solving Eq.(5).

We choose following parameters
for $V(\phi )$
\be m^2=30,\qquad \delta =26.0,\qquad \lambda =4\ee

 The choice of parameters is governed by the requirements that the bubble
size as well as its action  be small for appropriate values of the
temperature. The condition for high temperature approximation to be valid
is that $T >> r_0^{-1}$, where $r_0$ is the radius of the critical bubble in 3
dimensional Euclidean space. From now on we will use the Higgs mass
$m_H$ (= 8.37 for above choice of parameters) to define our mass
scale. The value of temperature we choose is $T = 0.6$ in these units.
For the above choice of parameters, $r_0^{-1}$ is found to be about 0.14
which justifies our use of high temperature approximation.

The solution of Eq. (5) is a bubble which remains static when evolved by
the classical equations of motion in the Minkowski space.
Expanding bubbles are the ones which are somewhat larger than this
bubble and we construct such a critical bubble by first
choosing $\delta = 25.9$
and finding the corresponding static bubble. This bubble when evolved
by the equations of motion with $\delta = 26.0$ becomes an expanding
bubble. Similarly, the subcritical bubble is found by finding the static
bubble with $\delta = 26.1$ which collapses when evolved with $\delta = 26.0$.

 As we have mentioned earlier, in the high temperature approximation our theory
effectively  becomes two dimensional. For a theory with one real
scalar field in two Euclidean dimensions the pre-exponential factor arising
in the nucleation rate of critical bubbles has been analytically calculated by
Voloshin [6]. We can therefore use results of [6] for our case as long as
we work within  high temperature approximation.  The pre-exponential
factor obtained from [6] for our case becomes

\be A =  {\epsilon\over 2\pi T}\ee

 It is important to note here that the results of [6] were for a single
real scalar field and one of the crucial ingredients used in [6]  for
calculating the pre-exponential factor was
the fact that for a bounce solution the only light modes contributing to
the determinant of fluctuations were the deformations
of the bubble perimeter. In the present case this is no longer true due to
the presence of the Goldstone boson which then also has to be accounted for
in the calculation of the determinant. We  will, however, not worry about
this in the present paper for the following reason. The nucleation rate
decides how frequent the bubble production is, and for our case with fixed
lattice size a moderate change in the nucleation rate simply amounts to a
change in the total duration of time in which all the bubbles are nucleated.
Thus as long as the nucleation rate is not changed by many orders of magnitude
the net effect is going to be a moderate change in the relative sizes of
various bubbles as they coalesce (since the time available to various bubbles
for expanding before coalescing will change). Of course, for large enough
change in the nucleation rate, even the net number of bubbles produced can
change which can then significantly effect our results of estimating the
probability of vortex formation. We will assume in this paper that the
inclusion of Goldstone bosons does not change the
nucleation rate by many orders of magnitude. We are investigating the
corrections induced by the Goldstone boson in the pre-exponential factor
of the nucleation rate.

  For the choice of parameters in Eq.(8), a plot of the potential is shown in
Fig. 1. We have added a constant to  $V(\phi )$
while plotting to make $V(\eta )=0$. The static bubble (solution of
Eq. (5)) has an outer radius of about 12.0 and is shown by the solid line in
Fig.2. The expanding critical bubble has radius $\simeq$ 12.1 (shown by the
dashed curve in Fig.2) while the subcritical bubble has radius $\simeq$ 11.9
(dotted curve in Fig.2). The radius of a bubble
is determined by the distance from the bubble center where  $\phi$ is very
small (appropriate to the lattice cutoff). On the other hand the
radius $r_0$
of the critical three dimensional bubble was determined by the distance
at which $\phi$ drops significantly (to about 1/e of the value of $\phi$
at the center of the
bubble). The values of the  nucleation rate $\Gamma$ (per unit volume per
unit time) for the static, critical and subcritical bubbles are respectively,
$1.98 \times 10^{-4}, ~ 1.67 \times 10^{-4}$ and $ 2.32 \times 10^{-4}$
(corresponding to temperature T = 0.6).

In our simulation critical bubbles represent the dominant class of expanding
bubbles,  and subcritical bubbles represent the dominant class of bubbles which
collapse. These bubbles are nucleated at any time
during the simulation with a uniform probability per unit time per unit volume
(governed by the respective nucleation rates as given above).
The location of the centers of the bubbles are also chosen at random.
If $\phi$ is very small in the region of interest (so that there is no bubble
there or in the immediate neighborhood), the false
vacuum is replaced by the bubble profile with a randomly chosen value of
the Higgs phase. Figs 3a-3b show nucleation of few scattered bubbles.
In plotting the Higgs phase,
the length of the arrows are chosen to be large for large $\phi$
and the direction of the arrow denotes $\theta$. Arrows are not plotted when
$\phi$ is very small.

After nucleation, bubbles are
evolved by time dependent equations of motion in the Minkowski space.

\be\Box \Phi_i =-{\partial V(\Phi )\over\partial\Phi_i},
                                             \qquad i=1,2. \ee

with ${\partial \Phi \over \partial t} = 0$ at t = 0. Here
$\Phi =\Phi_1+i\Phi_2$ and $\Box$ is the d'Alembertian operator in
2+1 dimensions.  As we have noted before, the critical
bubbles expand and coalesce with other bubbles while the subcritical bubbles
eventually collapse though they may oscillate for a while.

The bubble evolution was numerically implemented by a stabilized leapfrog
algorithm of second order accuracy both in space and in time with the
d'Alembertian operator  approximmated by a diamond shaped grid [3].
Lattice spacing in the spatial directions was chosen to be $\Delta x=.084$
and the spacing in the time direction was chosen to be $\Delta t=.059$. This
satisfies the criterion for the stability of the numerical evolution
that the Courant number C $\equiv {\Delta t \over \Delta x} \le {1 \over
\sqrt{d}}$, where $d$ is the number of spatial dimensions (2 in our case).
The simulations were performed on the Cray-2 supercomputer at
the Minnesota Supercomputer Institute.

Our simulation resulted in the production of $25$ critical and $43$ subcritical
bubbles, the ratio of their numbers being  about 0.58.
The expected ratio form their nucleation rates is
0.72. The last critical bubble was nucleated at $t = 19.97$ while the last
subcritical bubble was nucleated at $t = 41.05$.
We performed the simulation upto $t = 67.35$. Energy was well conserverd in
the early stages (to within about 4\% upto $t \simeq 47$). During the late
satges of evolution and well after the defects were formed, some of the
expanding bubbles partly escaped from the lattice leading to the
leakage of energy out of the region.
A snapshot of the process of bubble expansion and coalescence
is illustrated in Fig. 4a and in Fig. 4b. A plot of the Higgs phase
at the close of the simulation is shown in Fig. 4c.
Almost all the bubbles have collided by this time with
most of the region consisting of true vacuum and of vortices.

Vortices  are located by
looking for loops along which the phase of the Higgs field has a non-zero
winding number. We would like to emphasize that there
were no defects present at the start of
the simulation. Vortices and antivortices appeared in the the course of bubble
evolution if the randomly chosen Higgs phases of the colliding bubbles trapped
a nonzero winding number.
To illustrate these points in more detail, let us concentrate on a pair of
vortices shown in Fig. 5a which is a plot of $\eta - \phi$.
{}From the phase plot in Fig. 5b one can see that one of them is a vortex near
the coordinates $(x = 244, y = 119)$ and the other one is an antivortex near
$(x = 245, y = 98)$. These defects generally pick up
large speeds from unbalanced momenta of colliding  bubble walls at the time
of formation or due to asymmetric distribution of the Higgs phase, see [3].
Further, during their evolution they may be subjected
to random changes in momentum due to the energy emitted by the
decaying portions of walls. We plot the positions of the vortex and the
antivortex of Figs. 5a-5b respectively in
Figs. 6a-6b showing these effects. The average speed
for these vortices is $0.55$ (in c=1 units).

Let us look more closely at the vortex-antivortex pair shown in Figs. 5a-5b.
The vortex and antivortex should move closer due to
attractive forces and eventually annihilate each other. However, we can see
from Figs. 6a-6b that these objects are actually moving away from each other.
This is an interesting situation where the random velocities of the
vortex and antivortex are able to dominate ove the attractive forces between
them (which will be small due to large separation between  the vortex
and the antivortex).

Not all of the defects produced in our simulation are as clear as the ones
depicted in Fig. 5. Sometimes they are close to other defects and are difficult
to resolve. In other situations they are close to the false vacuum and we
discard them  on the grounds of stability. We find a total of $16$ defects
(two are connected to walls)
formed in our simulation over a region of size $335 \times 335$. As the number
of bubbles (critical and subcritical) is 68, this gives a probability of
vortex formation about 0.23 in
approximate agreement with the theoretical predictions of the vortex
formation probability (1/4 per correlation domain). We would like to mention
here that the number of vortices observed per bubble in [3] was larger than
the theoretical prediction in the early stages. [Though, in [3], several
vortex-antivortex pairs annihilated later leaving the final numbers
consistent with the theoretical predictions.]
However there is a crucial difference between the
simulation of [3] and the present work. In [3] all the bubbles were critical
bubbles whereas in the present simulation more than half of the bubbles are
subcritical bubbles which affect the vortex formation in a very different
manner. For example, a subcritical bubble either leads to only a short
lived vortex which eventually escapes into the false vacuum [3], or it leads
to the formation of a vortex-antivortex pair which annihilate each other,
see [7]. We mention that recently
an experimental investigation of the string formation
probability has been carried out in nematic liquid crystals with results
in good agreement with the theoretical predictions, see [8].

 We conclude by emphesizing the main aspects of our work. Here
we estimate the nucleation
rate for a given choice of parameters and nucleate critical as well as
subcritical bubbles with their respective nucleation probabilities. The result
is that most of the bubbles are subcritical bubbles which supress direct
collisions between critical bubbles. Therefore very few of the vortices
(only 3- 4) we see here are reasonably isolated and have clear prominent
profiles. These should be the ones produced by the collisions of critical
bubbles only. Most of the vortices are vortex-antivortex pairs with very small
separations and seem to be resulting from the collision of subcritical
bubbles with the critical ones. Counting all such vortices, the vortex
production is in agreement with the theoretical estimates.
However, the presence of subcritical bubbles seems to suppress the production
of prominent and clearly separated vortices (for 25 critical bubbles we
get only about 3-4 prominent vortices). One may expect that these are the only
ones which will eventually survive and the vortex-antivortex pairs which
are almost overlapping will all annihilate. Hence the defect production will
be suppressed due to the presence of subcritical bubbles.
We also estimate typical speed of such
objects imparted by various random processes operating during a phase
transition to be about .5 which is in agreement with the results in
earlier investigation [3].

 We are very grateful to M.B. Voloshin for very useful suggestions especially
regarding the estimation of the pre-exponential factor.
S.C. would like to thank Y.Hosotani for his encouragement while this work was
in progress.
This work is supported in part by the U. S. Department of Energy under contract
No. DE-AC02-83ER-40105 and by a grant from the Minnesota Supercomputer
Institute. The work of AMS has also been supported by the Theoretical Physics
Institute at the University of Minnesota and by the National Science Foundation
under Grant No. PHY89-04035.

\vskip.5in
\centerline{$\underline{\it REFERENCES}$}

\begin{enumerate}

\item T. W. B. Kibble, J. Phys. A9, 1387 (1976).

\item M.B. Voloshin, I.Yu. Kobzarev, and L.B. Okun, Yad. Fiz. 20, 1229(1974)
[Sov. J. Nucl. Phys. 20, 644(1975)]; S. Coleman, Phys. Rev. D15, 2929(1977).

\item A. M. Srivastava, Phys. Rev. D45, R3304(1992); Phys. Rev. D46,
1353 (1992).

\item M. Gleiser, E.W. Kolb and R. Watkins, Nucl. Phys. B364, 411 (1991).

\item A. D. Linde, Nucl. Phys. B216, 421 (1983).

\item M.B. Voloshin, Sov. J. Nucl. Phys. 42, 644 (1985).

\item A.M. Srivastava, University of Minnesota preprint, TPI-MINN-92/23-T,
(August, 1992).

\item M.J. Bowick, L. Chandar, E.A. Schiff and A.M. Srivastava,
Syracuse University preprint, SU-HEP-4241-512, TPI-MINN-92/35-T,
(August, 1992).

\end{enumerate}

\newpage

\centerline {\bf FIGURE CAPTION}

 Figure 1 : Plot of the potential. $V(\Phi)$ is plotted in the units of
$V(\eta)$ and $\phi$ in the units of $\eta$,
$\eta$ being the vaccum expectation value of $\Phi$. A constant
has been added to the potential so that $V(\eta)$ = 0.

 Figure 2 : Profiles of bubbles.
The solid curve shows the static bubble. The critical bubble is slightly
larger and is  shown by the dashed curve on the outside of the solid curve.
The subcritical bubble is slightly smaller than the static bubble and is
shown by the dotted curve on the inside  of the solid curve.
The length scale is in the Higgs mass units.

 Figure 3 : (a) shows the profiles of few scattered bubbles at t = 2.93.
The bubbles are randomly nucleated. (b) shows the Higgs phase plot for the
bubbles at the same stage.

 Figure 4 : (a) Plot of $\eta - \phi$ showing the coalescence of bubbles
at t = 41.00. (b) Higgs phase plot at the same stage. (c) Higgs phase
plot at the close of the simulation.

 Figure 5 : (a) Profiles of a vortex - antivortex pair. Note the deformed
configurations of $\phi$. (b) Higgs phase plot for the same pair.

 Figure 6 : (a) Trajectory of the vortex in Fig. 5
form $t=52.71$ to $t=67.35$.
 Arrowheads show the
locations of the vortex at succesive time steps $\Delta t =2.93$. (b) Same
for the antivortex of Fig. 5. Random motion of the pair is clear as is their
motion away from each other on the average. Average speed of all these vortices
is $\simeq 0.55c$.

\end{document}